\documentclass[12pt,preprint]{aastex}

\input{epshyf.sty}

\slugcomment{ApJ accepted}

\shorttitle{Two-Component Jets in GRBs}
\shortauthors{Huang et al.}

\begin{document}



\title{Rebrightening of XRF 030723: Further Evidence for a 
     Two-Component Jet in Gamma-Ray Burst}


\author{Y. F. Huang\altaffilmark{1,2}, X. F. Wu\altaffilmark{1},
Z. G. Dai\altaffilmark{1}, H. T. Ma\altaffilmark{1}, 
and T. Lu\altaffilmark{3,4} }






\altaffiltext{1}{Department of Astronomy, Nanjing University, Nanjing 210093, China
     (hyf@nju.edu.cn)}
\altaffiltext{2}{LCRHEA, IHEP, Chinese Academy of Sciences, Beijing 100039, China}
\altaffiltext{3}{Purple Mountain Observatory, Chinese Academy of Sciences, 
     Nanjing 210008, China}
\altaffiltext{4}{National Astronomical Observatories, Chinese Academy of Sciences, 
     Beijing 100012, China}


\begin{abstract}
Optical afterglows from two-component jets under various configurations
are investigated numerically. Generally, the light curve is 
characterized by a rapid rebrightening when the observer is off-axis
with respect to the narrow component, with the amplitude and 
peak time depending on detailed parameters. We further show that the
optical afterglow of XRF 030723, especially its notable and rapid rebrightening, 
can be well explained by a typical two-component jet. This X-ray 
flash, together with GRB 030329, strongly hints the two-component jet
model as a unified picture for X-ray flashes and gamma-ray bursts. 
With a narrow but ultra-relativistic inner outflow and a wide but
less energetic outer ejecta, a two-component jet will be observed 
as a typical gamma-ray burst if our line of sight is within the 
angular scope of the narrow outflow. Otherwise, if the line of sight
is within or slightly beyond the cone of the wide component, an 
X-ray flash will be detected.  
\end{abstract}




\section{Introduction}

Jets play an important role in gamma-ray bursts (GRBs) (Rhoads 
1997). Theoretically, collimation is essential to resolve the energy
crisis derived from a few GRBs, such as GRB 971214, 990123, and 
990510 (Bloom et al. 1998; Kulkarni et al. 1998, 1999; Andersen et 
al. 1999; Harrison et al. 1999). The existence of jets also sheds
light on the enigmatic nature of GRBs, indicating that the births 
of accreting, rapidly rotating black holes might be involved. Till 
now, jets in GRBs have been investigated in great detail, and 
many important effects are revealed, such as the break in afterglow
light curves (e.g., Rhoads 1999;  Sari, Piran, \& Halpern 1999; 
Panaitescu \& Kumar 2001; Huang \& Cheng 2003), orphan afterglows 
(e.g., Rhoads 1997; Granot et al. 2002; Huang, Dai, \& Lu 2002), and 
polarization (e.g., Gruzinov 1999; Hjorth et al. 1999; Mitra 2000; Rol 
et al. 2003; Coburn \& Boggs 2003; Waxman 2003). Recently it is 
further realized that collimated jets, when off-beamed, can also 
give birth to a special kind of GRBs, X-ray flashes (XRFs). 

XRFs have been identified as a sub-class of GRBs only recently. They
are characterized by relatively softer spectra, but with durations
and temporal structures typical of most long GRBs (Frontera et al. 
2000; Heise et al. 2003; Kippen et al. 2003; Barraud et al. 2003). 
However, in addition to the off-beam GRB mechanism (Woods \& Loeb 
1999; Nakamura 1999; Yamazaki, Ioka, \& Nakamura 2002, 2003; 
Jin \& Wei 2003), XRFs can also be produced by dirty fireballs 
(Dermer, Chiang, \& B\"ottcher  1999; Zhang \& M\'esz\'aros 2002b; 
Heise et al. 2003), or namely failed GRBs (Huang, Dai, \& Lu 2002). 
As the newly identified sub-class, XRFs still seem to be quite enigmatic, 
but can hopefully give useful hints on the nature of typical GRBs. 

In the field of X-ray rich GRBs, optical afterglows were first 
observed (Soderberg et al. 2002) from XRF 020903 (Sakamoto et al. 2003), 
with the redshift being measured as $z=0.251$ (Soderberg et al. 2003b). 
XRF 030723 (Prigozhin et al. 2003) is 
another important event, whose optical afterglow light curve has 
even been satisfactorily determined. It is quite striking that 
its afterglow decays in a way very similar to that of typical 
GRBs, i.e., following a simple power-law function of time. This 
strongly indicates that XRFs and other long GRBs may have similar 
origins. However, a prominent characteristics of the optical 
counterpart of XRF 030723 is that it rebrightened 
by about one magnitude from $t \sim 11$ d to $t \sim 14$ d 
(Fynbo et al. 2003b). Such a rapid rebrightening is somewhat 
unexpected to us. 

In this paper, we show that the behaviour of the optical afterglow 
of XRF 030723 can be well 
explained by the two-component jet model advocated by Berger et al. 
(2003b). Our results suggest that both the baryon loading mechanism 
and the off-beam mechanism are taking effect in XRF 030723. The 
structure of our paper is organized as follows. The model is described 
briefly in \S 2. We then numerically investigate the afterglow 
behaviour of two-component jets in \S 3, presenting optical light 
curves under various configurations. Our fit to the optical afterglow 
of XRF 030723 is illustrated in \S 4. \S 5 is the conclusion and 
discussion. 

\section{Model}

\subsection{Two-component jet}

The simplest jet model involves a homogeneous conical outflow. But in
reality a jet can be complicatedly structured (Zhang, Woosley, \& Heger 
2003a; Zhang, Woosley, \& MacFadyen 2003b). In this case, it is
usually assumed that the energy per unit solid angle depends as 
a power-law or a Gaussian function on the angular distance 
from the axis (M\'esz\'aros, Rees, \& Wijers 1998; Dai \& Gou 2001; 
Rossi, Lazzati, \& Rees 2002; Zhang \& M\'esz\'aros 2002a; 
Kumar \& Granot 2003; Salmonson 2003; Granot \& Kumar 2003; Zhang, 
Dai, Lloyd-Ronning, \& M\'esz\'aros 2003). Recently, our attention 
was drawn to another special kind of structured jets, i.e., 
two-component jets (Berger et al. 2003b; Sheth et al. 2003). 

A two-component jet has two components: a narrow ultra-relativistic 
outflow and a wide but mildly relativistic ejecta (Berger et al. 2003b;
Sheth et al. 2003; also see: Frail et al. 2000; Ramirez-Ruiz, 
Celotti, \& Rees 2002). At first glance, the two-component jet model 
still seems to be too coarse, but it strikingly gives 
a perfect explanation to the multi-band observations of GRB 
030329. As suggested by Berger et al. (2003b), the $\gamma$-ray and 
early afterglow emission of GRB 030329 come from the narrow 
component, while the radio and optical afterglow beyond 1.5 days 
should be produced by the wide component. They even derived the 
half opening angles of the two components as $\sim 5^{\rm o}$ and 
$\sim 17^{\rm o}$ respectively. The total intrinsic kinetic energy
of these two components is perfectly consistent with the standard 
energy reservoir hypothesis (Frail et al. 2001; Panaitescu \& 
Kumar 2001; Bloom, Frail, \& Kulkarni 2003; Berger, Kulkarni, \& Frail 2003a). 

In this paper, we designate the initial half opening angle of the 
narrow and the wide component as $\theta_{\rm 0,N}$ and 
$\theta_{\rm 0,W}$, respectively, where the subscript ``N'' means
``narrow'' and ``W'' means ``wide''. We further assume that 
their isotropic equivalent kinetic energies are $E_{\rm N,iso}$ 
and $E_{\rm W,iso}$, and initial Lorentz factors are 
$\gamma_{\rm 0,N}$ and $\gamma_{\rm 0,W}$, respectively. We will 
model the dynamical evolution and radiation process of the two
components in \S 2.2, and calculate their optical afterglows 
numerically in \S 3. 

\subsection{Dynamics and radiation process}

We assume that the two components are coaxial. To simplify the 
problem, the interaction and overlapping of the two components 
are generally neglected. We then can essentially calculate the dynamical 
evolution of the two components independently, and add their 
emission together to get the total afterglow light curve of 
the entire jet. 

We use the model developed by Huang et al. (Huang, Dai, \& Lu 
2000a; Huang et al. 2000b) to describe
the evolution and radiation of each component. In this model, 
evolution of the bulk Lorentz factor is given by (Huang, Dai, \& 
Lu 1999a, b), 
\begin{equation}
\label{dgdm1}
\frac{d \gamma}{d m} = - \frac{\gamma^2 - 1}
       {M_{\rm ej} + \epsilon m + 2 ( 1 - \epsilon) \gamma m}, 
\end{equation}
where $m$ is the mass of swept-up interstellar medium (ISM), $M_{\rm ej}$ 
is the initial mass of the ejecta, and $\epsilon$ is the radiative 
efficiency. Equation~(1) has the virtue of being applicable in both 
the ultra-relativistic and the non-relativistic phases (Huang et al. 
1999a, b). The lateral expansion of the outflow is described realistically
by (Huang et al. 2000a, b),
\begin{equation}
\label{dthdt2}
\frac{d \theta}{d t} = \frac{c_{\rm s} (\gamma + \sqrt{\gamma^2 - 1})}{R},
\end{equation}
with
\begin{equation}
\label{cs3}
c_{\rm s}^2 = \hat{\gamma} (\hat{\gamma} - 1) (\gamma - 1) 
	      \frac{1}{1 + \hat{\gamma}(\gamma - 1)} c^2 , 
\end{equation}
where $\theta$ is the half opening angle, $c_{\rm s}$ is the co-moving 
sound speed, $R$ is the radius, and  
$\hat{\gamma} \approx (4 \gamma + 1)/(3 \gamma)$ is the adiabatic 
index. For simplicity, we assume that the ejecta is adiabatic, which
means the radiative efficiency in equation~(1) is $\epsilon \equiv 0$.  

Optical afterglows can be calculated by considering synchrotron 
radiation from shock-accelerated electrons in the outflow. In our model, 
we use a realistic electron distribution function that takes into account
the cooling effect (Sari, Piran, \& Narayan 1998) and the non-relativistic
effect (i.e., the minimum Lorentz factor of electrons, $\gamma_{\rm e,min}$, 
will be less than a few when the outflow enters the ``deep Newtonian phase'' 
(Huang \& Cheng 2003)). Additionally, the equal arrival time surface 
effect (Waxman 1997; Sari 1998; Panaitescu \& M\'esz\'aros 1998) 
is also included in our consideration. 

\section{Numerical Results}

We now present our numerical results for afterglows of two-component 
jets. For convenience, let us first define a set of ``standard'' parameters
as following: $\theta_{\rm 0,W} =0.3$, $E_{\rm W,iso}=10^{52}$ ergs, 
$\gamma_{\rm 0,W}=30$, $\theta_{\rm 0,N}=0.1$, $E_{\rm N,iso}=5
\times 10^{53}$ ergs, $\gamma_{\rm 0,N}=300$, ISM number density 
$n=1$ cm$^{-3}$, electron energy ratio $\epsilon_{\rm e}=0.1$, 
magnetic energy ratio $\epsilon_{\rm B} = 0.01$, luminosity distance
$D_{\rm L}=2$ Gpc, electron distribution index $p=2.5$, and viewing 
angle $\theta_{\rm obs}= 0.35$. These parameter values are typical 
in GRB afterglows. Most of them are also roughly consistent with those derived for 
GRB 030329 by Berger et al. (2003b). A two-component jet with this standard
configuration has an intrinsic kinetic energy of $\sim 1.5 \times 10^{51}$
ergs, also approximately meets the requirement of the standard energy reservoir
hypothesis (Frail et al. 2001; Panaitescu \& Kumar 2001; Bloom et al. 2003; 
Berger et al. 2003a). 

Figure 1 illustrates R-band afterglows from a two-component jet under 
the ``standard'' configuration. It also shows clearly how emission from 
the two components is compounded together to give out the total light 
curve. In the current configuration, since the line of sight is notably 
outside the initial angular range of the narrow component, we see that
the afterglow is dominated by emission from the wide component before 
$\sim 10^6$ s. But as the narrow component decelerates and expands 
laterally, its emission increases rapidly until the flux density finally
comes to a peak at about $\sim 1.6 \times 10^6$ s. The total light curve 
is dominated by this component thereafter. 

At the peak time mentioned above, the Lorentz factor of the narrow 
component is $\gamma_{\rm N} \approx 2.4$, and its angular width is 
$\theta_{\rm N} \approx 0.24$. Note that $\theta_{\rm N}$ is 
still notably less than $\theta_{\rm obs}$, but the inverse of the 
Lorentz factor ($1/\gamma_{\rm N} \approx 0.4$) is very close to
$\theta_{\rm obs}$. We conjecture that the peak time is roughly 
determined by the condition of $\gamma_{\rm N} \sim 1/\theta_{\rm obs}$. 
This conjecture is confirmed by our further calculations under 
other configurations. 

Another important characteristics that deserves mentioning in Figure 1
is the rapidness of the increase of the narrow component emission 
before the peak point. This behaviour hints that a two-component jet can 
potentially explain the quick rebrightening of optical afterglows of 
XRF 030723. 

The shape of the total optical light curve will surely be affected by the 
energy ratio of the two components. To illustrate the effect, we fix the 
isotropic equivalent energy of the wide component, but vary that of the
narrow component, and calculate the afterglow emission. The results are 
presented in Figure~2. Generally speaking, a more powerful narrow 
component can reasonably result in a more prominent peak, but with the
peak time slightly postponed. From Figure~2, we can also imagine that in 
some special cases, when $E_{\rm N,iso}$ is not much larger than 
$E_{\rm W,iso}$, the peak will become anonymous so that we can observe
nothing but a short plateau in the total light curve. 

Another factor that affects the energy ratio of the two components is 
their relative angular width. In Figure~3, we illustrate the impact of 
$\theta_{\rm 0,N}$ on the total light curve. Again, we see that a larger
$\theta_{\rm 0,N}$, which leads to a more powerful narrow component, makes
the peak more prominent. However, unlike the effect of $E_{\rm N,iso}$ 
(see Figure~2), a larger $\theta_{\rm 0,N}$ tends to make the light curve
peak slightly earlier. This in fact is not difficult to understand.

Figure 4 illustrates the effect of viewing angle ($\theta_{\rm obs}$) on 
the light curve. When $\theta_{\rm obs} \leq \theta_{\rm 0,N}$, the entire
light curve is dominated by the narrow component emission. When 
$\theta_{\rm obs} > \theta_{\rm 0,N}$, the light curve is dominated by 
the wide component emission initially, but is dominated by emission from 
the narrow component at late stages. It is clear that the viewing angle 
affects the peak time notably. It should also be noted that the light curve 
of $\theta_{\rm obs}=0.1$ differs from that of $\theta_{\rm obs}=0$ only 
slightly. Similar behaviour can also be observed in the light curves 
of $\theta_{\rm obs}=0.2$ and $\theta_{\rm obs}=0.3$ at early stages.  
It means the light curves will almost be the same as long as
the line of sight is within the initial angular range of the outflow, 
consistent with previous studies 
(Huang et al. 2000a). However, the light curve difference between 
$\theta_{\rm obs}=0.2$ and $\theta_{\rm obs}=0.3$ at early phase is 
noticeably larger than that between the $\theta_{\rm obs}=0$ and 
$\theta_{\rm obs}=0.1$ curves. This is because the wide component
is expanding with a much smaller Lorentz factor ($\gamma_{\rm 0,W}
=30$). 

In Berger et al.'s study of GRB 030329, the observer is assumed to 
be on the axis of the two-component jet (Berger et al. 2003b). The 
overall R-band optical afterglow can then be divided into three 
different stages. The early optical afterglow ($t \leq 1$ d), which
shows a notable light curve break (evolving from $\sim t^{-1}$ to 
$\sim t^{-2.3}$) at about $t \sim 0.6$ d, is believed to originate 
from the narrow component; The mid-term afterglow (1.5 d $\leq t <$ 
20 d) comes from the wide component; Optical emission at late 
stages ($t > 20$ d) is attributed to a hidden supernova. Of special
interest is the fact that in their calculation, the emission is 
dominated by the narrow component in the first stage (i.e. $t \leq 1$
d), and then dominated by the wide component thereafter (i.e. 
1.5 d $\leq t <$ 20 d). This seems to be completely opposite to the 
cases illustrated in our Figures 1 --- 4, where the emission is 
generally dominated by the wide component initially, and then by the
narrow component at late times. Note that in most of our calculations, 
$\theta_{\rm obs}$ has been taken as 0.2 or even larger, so that the 
observer is initially well outside the opening angle of the narrow jet. 
The reversal of the roles played by the wide and narrow components is 
thus in fact due to the difference of the viewing angle, and to the 
difference in the relative intrinsic kinetic energies of the two 
components as well, as explained further below. 

In Figure 4 we have also evaluated $\theta_{\rm obs}$ as 
zero in one case, i.e. the thick 
dashed line. In this case, the light curve is really dominated by 
the narrow component initially, but a bump or a plateau indicating 
the contribution from the wide component can barely be seen even at 
late stages. In other words, the light curve is virtually dominated 
by the narrow component all the time. One thus may doubt the ability
of the two-component jet model to reproduce the optical light curve 
of GRB 030329 as suggested by Berger et al. In fact, there are slight
differences in the parameter values taken by Berger et al. and by us. 
Berger et al. (2003b) evaluated their parameters as  
$\theta_{\rm 0,W} \sim 0.3$, $E_{\rm W,iso} \sim 5.6 \times 10^{51}$ ergs, 
$\theta_{\rm 0,N} \sim 0.09$, and $E_{\rm N,iso} \sim 1.2 \times 10^{52}$ ergs. 
The intrinsic kinetic energies of their wide and narrow components 
are then $\sim 1.2 \times 10^{50}$ ergs and $\sim 2.4 \times 10^{49}$ ergs 
respectively. Note that their $E_{\rm N,iso}$ is only twice their 
$E_{\rm W,iso}$, so that the intrinsic kinetic energy of the wide component
is essentially larger than that of the narrow component by a factor of $\sim 5$. 
It is then easy to understand that the late time emission should be dominated by
the wide component. But in our ``standard'' configuration, $E_{\rm N,iso}$ 
is higher than $E_{\rm W,iso}$ by a factor of $\sim 50$, and the intrinsic kinetic 
energies of the wide and narrow components are $\sim 2.2 \times 10^{50}$ 
ergs and $\sim 1.2 \times 10^{51}$ ergs respectively. We see that the 
intrinsic kinetic energy of our narrow compoent is virtually $\sim 6$ times 
as large as that of our wide compoent. This explains why the thick dashed 
line in our Figure 4 is overall dominated by the narrow component. 

Anyway, it should be of some interest to examine numerically whether the 
two-component jet model is really capable of reproducing the optical 
afterglow of GRB 030329 or not. We have repeated our calculation by taking 
a set of parameters that are very close to those recommended by Berger 
et al. (2003b), i.e. 
$\theta_{\rm 0,W} = 0.3$, $E_{\rm W,iso} = 5 \times 10^{51}$ ergs, 
$\theta_{\rm 0,N} = 0.08$, $E_{\rm N,iso} = 10^{52}$ ergs, 
and $n = 2$ cm$^{-3}$. The results are presented in Figure 5. In Figure 5a we
see that the wide component emission (dashed line) is really stronger than 
the narrow component emission (dash-dotted line) after $\sim 4 \times 
10^4$ s. The slope of the dashed line is also obviously flatter than that 
of the dash-dotted line. A two-component jet thus does have the potential 
of reproducing the behavior of GRB 030329. However, Figure 5a also clearly 
shows that the total light curve differs from that of GRB 030329 markedly. 
The major problem is that although the wide component emission is weaker than 
the narrow component emission before $\sim 4 \times 10^4$ s, it is weaker only
slightly. As a result, we cannot observe the expected flatening after 
$\sim 4 \times 10^4$ s, nor can we observe the jet break caused by the 
narrow component before $\sim 4 \times 10^4$ s. 

Figure 5a indicates that while the two-component jet model basically has the 
potential of explaining the afterglow behavior of GRB 030329, the problem might 
in fact be much more complicate than what we have expected.  Maybe we need to 
consider many other details that may take effect in the process. In Berger et
al.'s (2003b) modeling of GRB 030329, the wide component emission is assumed to
be very weak before $\sim 2$ d, with the reason unexplained explicitly. We should
bear in mind that our consideration of the two-component jet model here is in
fact still very preliminary. Many details, some of which we believe do have the effect
of reducing the early emission from the wide component, have been omitted for 
simplicity. For example, we do not consider the overlap of the two components. 
In fact, since there is a central narrow component resided on the axis, the
wide component should be empty at the center in reality. In other words, 
the wide component should be a hollow one. It should be invisible to the 
observer at early times as long as the line of sight is within the central hole, 
i.e. $\theta_{\rm obs} < \theta_{\rm 0,N}$. In Figure 5b, we have re-drawn 
the light curves by cutting the emission from the central cone of the wide 
component occupied by the narrow one. We see that the wide component emission 
is really reduced markedly before $\sim 10^5$ s, and the total light curve 
now seems much better. Of course, this total light curve still cannot be 
entitled ``completely satisfactory'' as compared with the observed optical 
afterglow light curve of GRB 030329, but we believe a satisfactory fit 
should be possible when more details, including more realistic interaction
between the two components, are considered. We will not attempt to present
an overall fit to the multi-band observations of afterglows of GRB 030329 
here. It is beyond the scope of this study. 

\section{Fit to XRF 030723}

XRF 030723 occurred at 06:28:17.45 UT on July 23, 2003, and was 
detected by the High Energy Transient Explorer-II (HETE-2) 
(Prigozhin et al. 2003). The burst duration ($T_{90}$) in the 
7 --- 30 keV band was $\sim 25$ s, with a total fluence of 
$\sim 2 \times 10^{-7}$ ergs cm$^{-2}$. In contrast, its fluence
in 30 --- 400 keV  band was less than $7 \times 10^{-8}$ ergs 
cm$^{-2}$, which is less than 0.4 times the fluence in the 
7 --- 30 keV band. An XRF nature thus is strongly favored 
by these HETE-2 observations (Prigozhin et al. 2003). 

A preliminary localization with an error radius of $\sim 30'$ was 
distributed through the internet as early as 42 s later, triggering 
an extensive campaign searching for its afterglows in various bands. 
The optical counterpart was first reported by Fox et al. (2003), and
then followed by many other groups. The transient has also been 
detected in X-rays (Butler et al. 2003a, b), but no radio counterpart 
was observed with a $3 \sigma$ upper limit of 180 $\mu$Jy at 
8.46 GHz on July 26.42 UT (Soderberg et al. 2003a). 

XRF 030723 is an important event in the studies of XRFs. For the 
first time the optical afterglow light curve is determined  
satisfactorily for an XRF. Strikingly but not unexpectedly, the 
optical transient decays in a way very similar to that of typical 
GRBs, i.e., roughly following a power-law of time, strongly 
indicating that XRFs and other long GRBs may have similar origins. 

However, a special feature of XRF 030723 is that the optical transient
rebrightened by about one magnitude from $t \sim 11$ d to $t \sim 
14$ d (Fynbo et al. 2003b). Such a rapid increase in the afterglow
light curve (see Figure~6) needs an appropriate explanation.
In a recent study, Dado et al. (Dado, Dar, \& De R\'ujula 2003) 
have interpreted the afterglow of XRF 030723 in the frame-work of 
their cannonball model. They attributed the rebrightening to a 
hidden supernova (also see Fynbo et al. 2003b). However, we 
notice that a typical type Ib/c supernova usually cannot enhance 
the emission so steeply. 

From Figures 1 --- 4 we have seen that the rapid rebrightening 
is a general feature for two-component jets if the observer is 
off-center. Motivated by this phenomenon, we have tried to use 
a two-component jet to model the R-band light curve of XRF 030723. 
In Figure~6, we illustrate our best fit, where the parameters are 
taken as: $\theta_{\rm 0,W} =0.3$, $E_{\rm W,iso}=10^{52}$ ergs, 
$\gamma_{\rm 0,W}=30$, $\theta_{\rm 0,N}=0.09$, $E_{\rm N,iso}= 3
\times 10^{53}$ ergs, $\gamma_{\rm 0,N}=300$,  $n=1$ cm$^{-3}$, 
$\epsilon_{\rm e}=0.1$, $\epsilon_{\rm B} = 0.01$, $p=3.2$, 
and $\theta_{\rm obs}= 0.37$. Under this configuration, the total 
intrinsic kinetic energy is $\sim 8.3 \times 10^{50}$ ergs, 
consistent with the standard energy reservoir hypothesis. Since the 
redshift is still unknown, we arbitrarily take the luminosity 
distance as $D_{\rm L}=2.5$ Gpc.  In our calculation, we omit 
the overlapping of the two components since the observer is 
outside the scope of the narrow component. 
Note that the error bars of the two 
observational data points near $10^6$ s are not plotted,
because they are not available in the literature. But we can imagine that 
they might be comparable to those of the points which are just 
preceding or behind them. Figure~6 shows clearly that the two-component
jet model can reproduce the light curve satisfactorily. 

In our fit, we have taken $\theta_{\rm obs}$ as 0.37. If we change 
$\theta_{\rm obs}$ to 0.35 or an even smaller value, then the 
theoretical light curve will be notably above those upper limits 
and flux densities (totally 6 points) obtained by ROTSE-III between 
$t \sim 50$ s and $t \sim 2600$ s (Smith et al. 2003). This is a 
good example showing how the early afterglows can provide valuable 
clues to our understanding of the nature of GRBs. 

Our results strongly suggest that XRF 030723 might be produced by
a two-component jet. The central narrow component has an initial 
Lorentz factor typical of most GRB fireballs (i.e., $\sim 100$ 
--- 1000). It can generate a GRB successfully if viewed on or 
near the axis. But since the observer is highly off-center,  
the event is finally completely undetectable. The wide 
component has an initial Lorentz factor of $1 \ll \gamma_{\rm 0,W}
\ll 100$. It fails to produce a typical GRB, but can give birth
to an XRF (Huang et al. 2002). This XRF is detectable even though
our line of sight is slightly outside the angular range of the 
wide component. 

It should be noted that in our model, the main burst of XRF 030723 
is essentially attributed to the dirty fireball effect (Dermer et al.
1999; Huang et al. 2002), not to the off-beam effect. The off-beam
mechanism is included here mainly to meet the requirement of the 
early afterglow observations. 

\section{Conclusion and Discussion}

Optical afterglows from two-component jets under various configurations 
have been investigated numerically. The light curve is generally
characterized by a rapid rebrightening when the observer is off-beamed 
with respect to the narrow component. Depending on parameters such as 
$\theta_{\rm 0,W}, \theta_{\rm 0,N}, E_{\rm W,iso}, E_{\rm N,iso}$ and
$\theta_{\rm obs}$, the amplitude and peak time of the rebrightening 
vary accordingly. In some special cases, when the central component is 
not much more powerful than the outer, wide component, the rebrightening
will become so shallow that it simply manifests as a short plateau in the
light curve. In fact, when taking the parameters recommended by Berger 
et al. (2003b) for GRB 030329, we can imagine that no rebrightening 
or plateau would be seen in the total optical light curve even a
non-zero viewing angle was assumed (cf. Figure 5). In this case, since the 
intrinsic kinetic energy of the narrow component is much less than that of 
the wide one, we virtually would only see a simple light curve which barely 
differs from that produced by the wide component alone. 

XRF 030723 may play an important role in the studies of XRFs as well as GRBs. 
We have shown that its afterglow, especially the rapid rebrightening at 
$t \sim 14$ d, can be well presented by the two-component jet model. XRF 
030723, together with GRB 030329 (Berger et al. 2003b), strongly hints the 
two-component jet model as a unified picture for XRFs and GRBs. In this
frame-work, the central, narrow component has a Lorentz factor of 
$\sim 100$ --- 1000, and the outer, wide component has a Lorentz factor 
of $1 \ll \gamma_{\rm 0,W} \ll 100$ due to baryon loading. If the observer
is within the angular range of the central component, a typical GRB will
be observed. Otherwise, if the observer is within or slightly beyond the
scale of the wide component, an XRF or an X-ray rich GRB will be detected. 
Additionally, since the wide component is a highly decelerated outflow 
and has been affected seriously by circum-engine baryons, we can imagine
that it will lose its memory of the central engine so that the XRF emission
should be produced mainly by external shocks (Zhang et al. 2003b). 
Of course, strong variability is still possible in the XRF light curve, 
since the condition of the wide component as well as its environment may 
be very complicated in reality. 

Recently it has been realized that the homogeneous jet assumption is only 
a coarse approach to the outflows in GRBs. Structured jets, which find 
support in numerical simulations of massive star collapse (Zhang et al. 
2003a, b), are receiving more and more attention. 
In a recent study, Zhang et al. (2003a)
found that when an ultra-relativistic jet breaks out from a massive Wolf-Rayet
star, it is usually surrounded by a cocoon of less energetic, but still 
highly relativistic ejecta (see Ramirez-Ruiz, Celotti, \& Rees (2002) for
a theoretical understanding of the formation of the cocoon). 
Thus although the two-component jet model still 
seems to be too coarse, it in fact is supported by numerical simulations. 
It is very interesting that Zhang et al. (2003a) also noted that these
cocoons may give birth to XRFs. 
 
The rapid rebrightening of XRF 030723 should not be due to a hidden 
supernova. We have interpreted the rebrightening as evidence for the 
existence of a narrow component in a wider ejecta. But it deserves 
mentioning that a density-jump, i.e., an abrupt increase in the ISM 
density, can also give birth to a rapid rebrightening (Dai \& Lu 2002;
Lazzati et al. 2002; Dai \& Wu 2003). 
In this case, we expect that a simultaneous 
decrease in radio brightness could be seen since self-absorption 
will be enhanced in a denser environment. Lacking an ideal coverage
and multi-band observations between $t \sim 11$ d and $t \sim 14$ d, the 
density-jump interpretation for XRF 030723 cannot be completely excluded
currently. Still another possible interpretation may involve refreshed 
shocks, i.e. shells produced by the central engine at late times. This
mechanism has been adopted to explain the variability in the optical 
afterglows of GRB 021004 and 030329 (Nakar, Piran, \& Granot 2003; 
Granot, Nakar, \& Piran 2003). Refreshed shocks can potentially 
give birth to many rebrightenings in a single afterglow light curve. 
In the future, when afterglows are observed from more and 
more XRFs, these interpretations can then be tested in more details. 

Finally, since two-component jets are most likely produced in massive
star collapses, it is possible that XRF 030723 may be accompanied by 
a supernova. Wu et al. (2003) pointed out that the supernova 
component usually peaks at about $\sim (1+z) \times 15$ d in the optical
light curve of typical GRBs, where 15 d is approximately the peak time
of a type Ic supernova in its comoving frame. Since the redshift of 
XRF 030723 is not large ($z < 2.1$, Fynbo et al. 2003a), we expect 
that the supernova component should appear $\sim 20$ --- 40 days after
the trigger. A search for such a supernova component should deserve 
trying, because XRF 030723 may not be unacceptably far away. It is 
interesting that XRF 030723 has been observed at $t \sim 11$ d, 12 d,
64 d, and 71 d respectively. These observations show that the optical
transient is still decaying rapidly about 70 days later, further 
mornitoring of the optical transient is thus interesting and valuable. 
However, lacking observational data between $\sim 12$ d and 60 d, 
the existence of a supernova is unfortunately uncertain.

\acknowledgments

We thank an anonymous referee for valuable comments and suggestions 
that lead to an overall improvement of this study. 
This research was supported by the Special Funds for Major State
Basic Research Projects, the National Natural Science Foundation
of China (Grants 10003001, 10233010, and 10221001), the Foundation 
for the Author of National Excellent Doctoral Dissertation of P. R. 
China (Project No: 200125), and the
National 973 Project (NKBRSF G19990754).




\clearpage


\begin{figure} \centering 
\epsfig{file=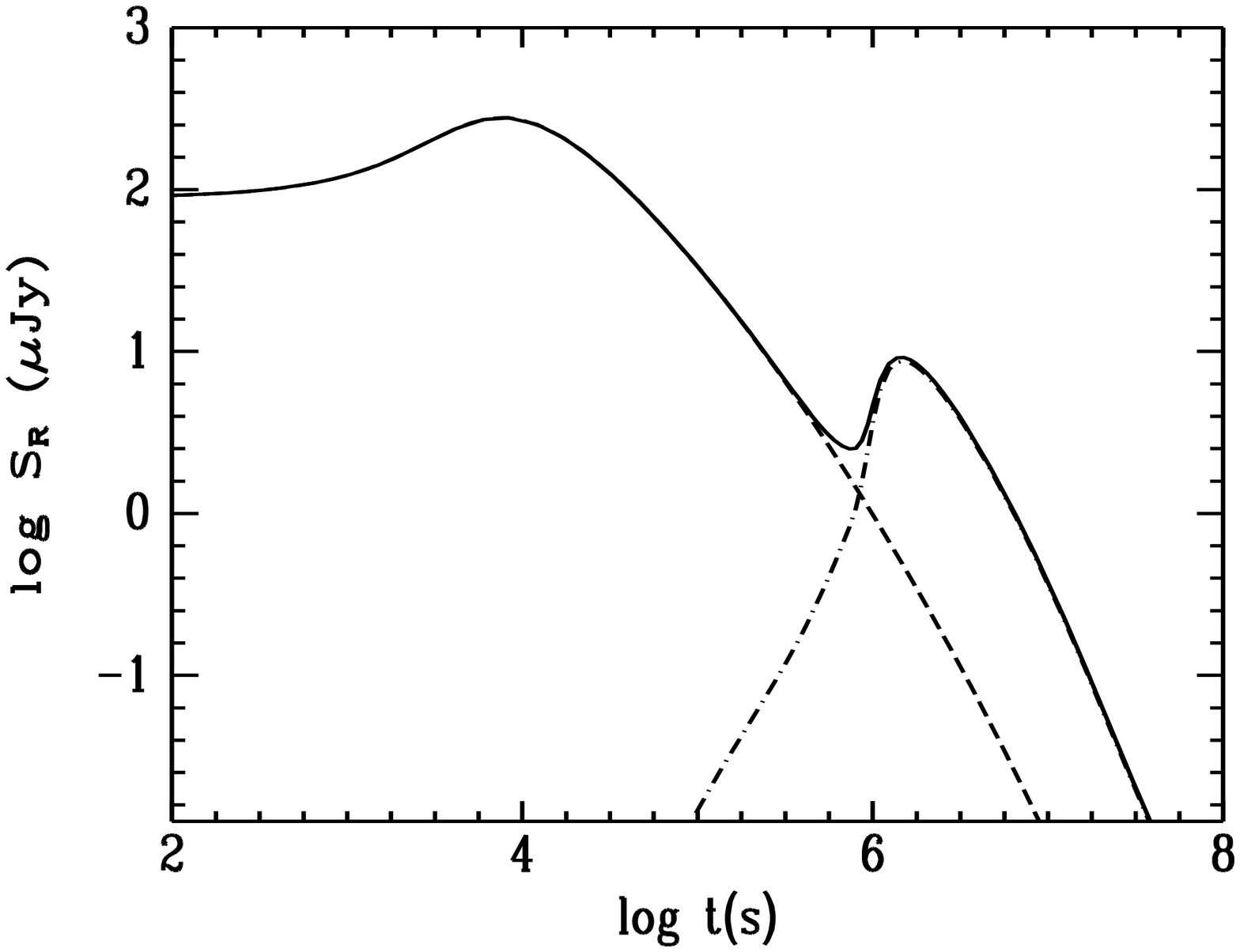, angle=-90, height=80mm, width=9cm, 
bbllx=120pt, bblly=125pt, bburx=530pt, bbury=575pt}
\caption{Optical (R-band) afterglow of a two-component jet with ``standard''
 parameters, which are defined in section 3 of the main text. The dashed line 
 corresponds to emission from the wide component and the dash-dotted line is 
 for the narrow component. The solid line illustrates the total light curve.   } 
\label{fig1}
\end{figure}

\begin{figure} \centering 
\epsfig{file=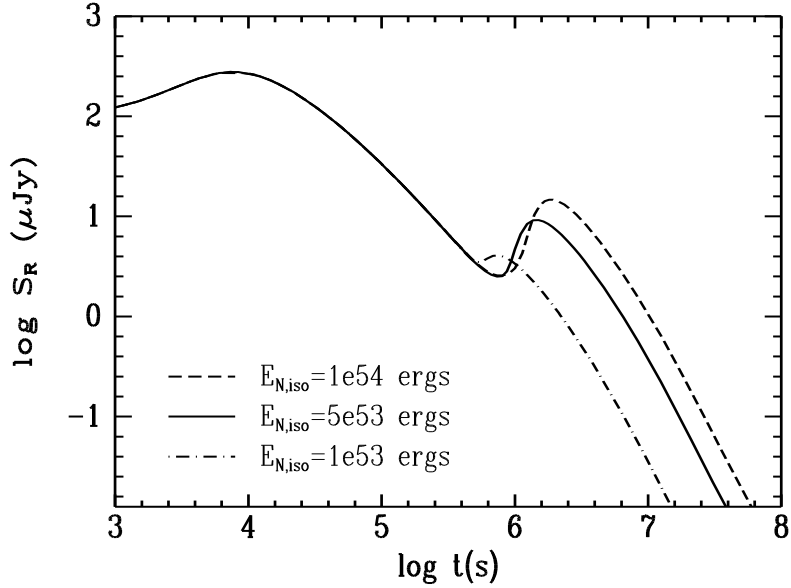, angle=-90, height=80mm, width=9cm, 
bbllx=120pt, bblly=125pt, bburx=530pt, bbury=575pt}
\caption{Effects of isotropic equivalent energy of the narrow component on 
  the afterglow from a two-component jet. The solid line is plot with 
  ``standard'' parameters defined in section 3. Other lines are drawn 
  with only $E_{\rm N,iso}$ altered.  } 
\label{fig2}
\end{figure}

\begin{figure} \centering 
\epsfig{file=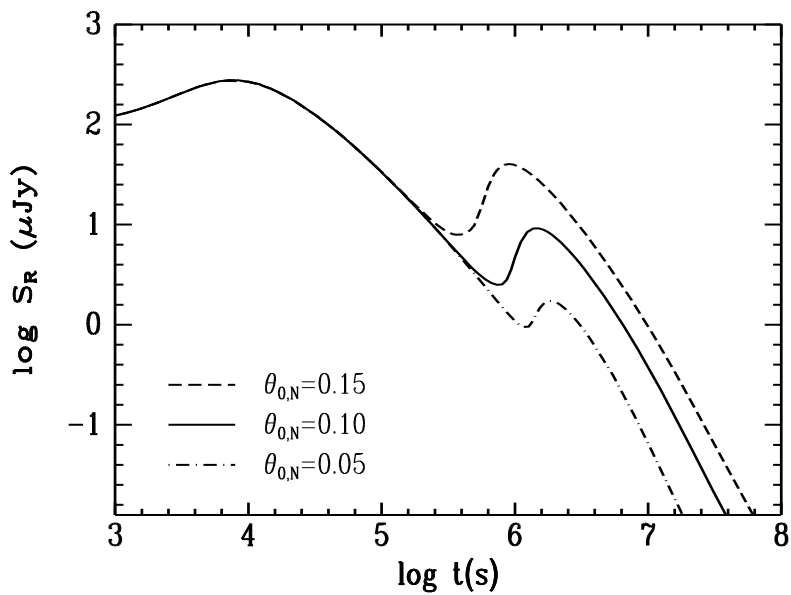, angle=-90, height=80mm, width=9cm, 
bbllx=120pt, bblly=125pt, bburx=530pt, bbury=575pt}
\caption{Effects of initial opening angle of the narrow component on the 
  afterglow from a two-component jet. The solid line is plot with 
  ``standard'' parameters defined in section 3. Other lines are drawn with
  only $\theta_{\rm 0,N}$ altered.    } 
\label{fig3}
\end{figure}

\begin{figure} \centering 
\epsfig{file=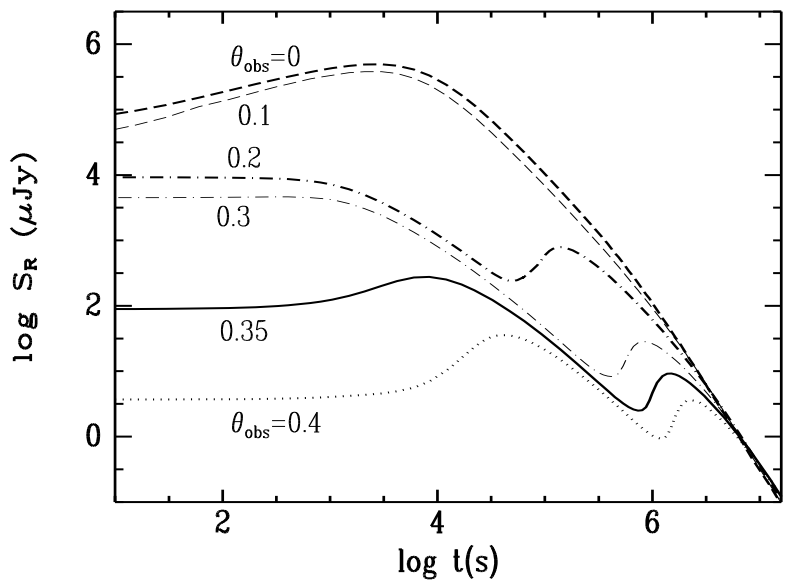, angle=-90, height=80mm, width=9cm, 
bbllx=120pt, bblly=125pt, bburx=530pt, bbury=575pt}
\caption{Effects of viewing angle on the afterglow from a two-component 
  jet. The solid line is plot with ``standard'' parameters defined in section
  3. Other lines are drawn with only $\theta_{\rm obs}$ altered.    } 
\label{fig4}
\end{figure}

\begin{figure} \centering 
\epsfig{file=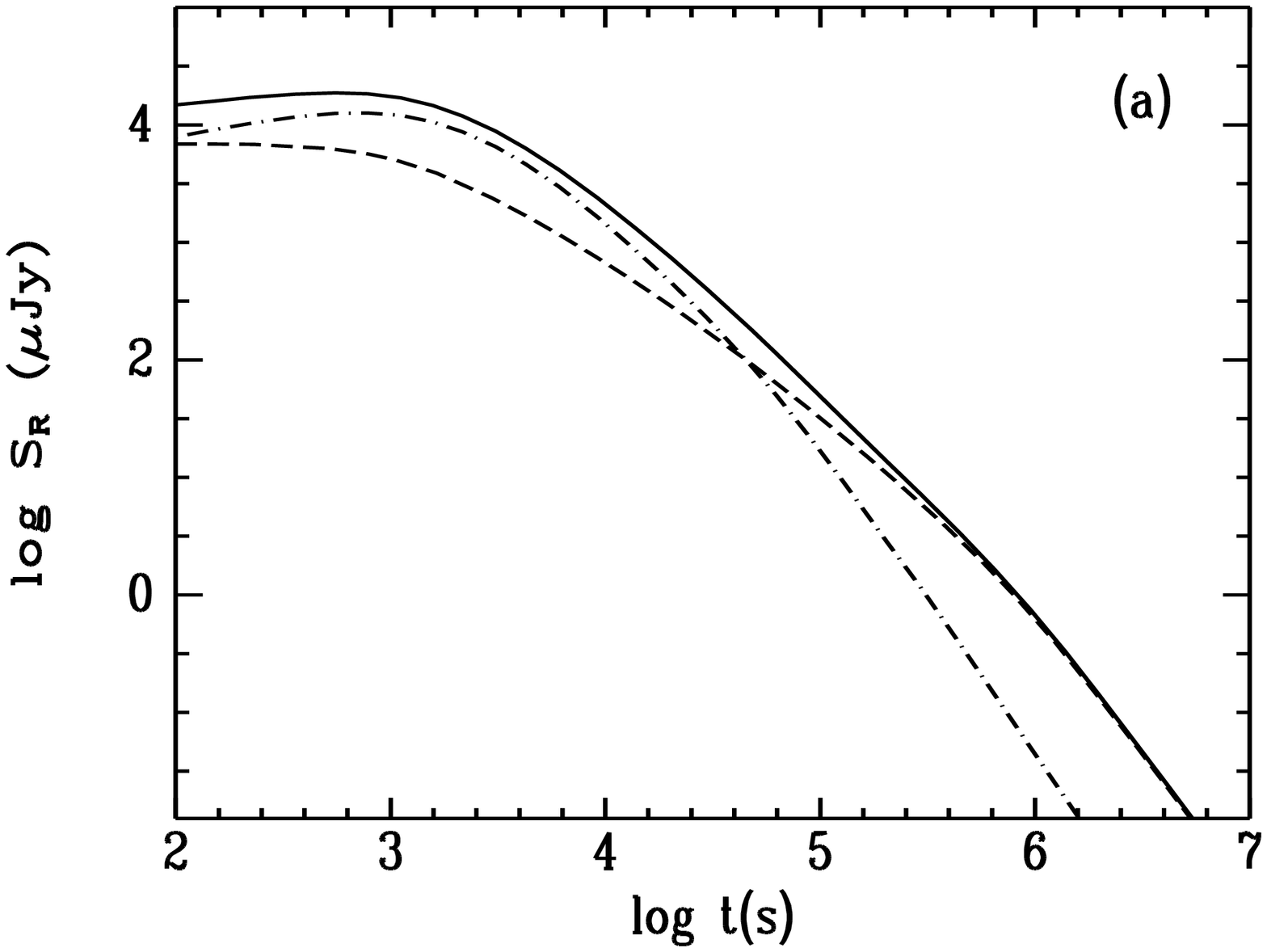, angle=-90, height=80mm, width=9cm, 
bbllx=120pt, bblly=125pt, bburx=530pt, bbury=575pt}
\epsfig{file=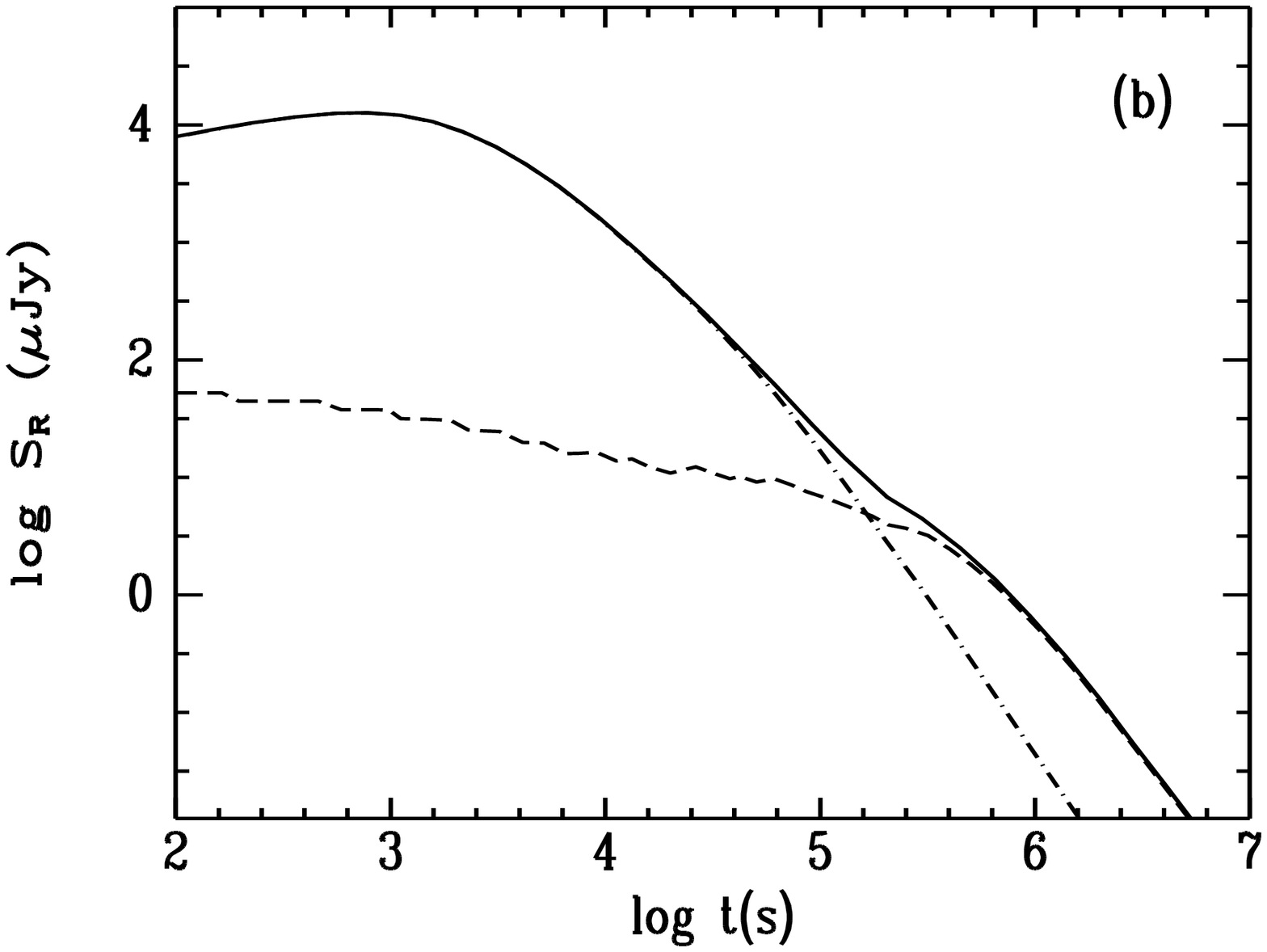, angle=-90, height=80mm, width=9cm, 
bbllx=120pt, bblly=125pt, bburx=530pt, bbury=575pt}
\caption{{\bf (a)} Afterglow from a two component jet when $E_{\rm N,iso}$ is only 
twice $E_{\rm W,iso}$ and when the observer is on-axis. Note that the 
parameters are evaluated as  
$\theta_{\rm 0,W} = 0.3$, $E_{\rm W,iso} = 5 \times 10^{51}$ ergs, 
$\theta_{\rm 0,N} = 0.08$, $E_{\rm N,iso} = 10^{52}$ ergs, 
$n = 2$ cm$^{-3}$, $\theta_{\rm obs}=0$, with other parameters 
the same as in Figure 1. Under this configuration, the wide component
has an intrinsic kinetic energy of $1.1 \times 10^{50}$ ergs, and the 
narrow compoent's energy is $1.6 \times 10^{49}$ ergs. 
The dashed line corresponds to emission 
from the wide component, the dash-dotted line corresponds to emission 
from the narrow component, and the solid line is the total light curve. 
{\bf (b)} Same as (a), except that the wide component is now assumed to be a 
hollow cone since its central portion is occupied by the narrow component.      } 
\label{fig5}
\end{figure}

\begin{figure} \centering 
\epsfig{file=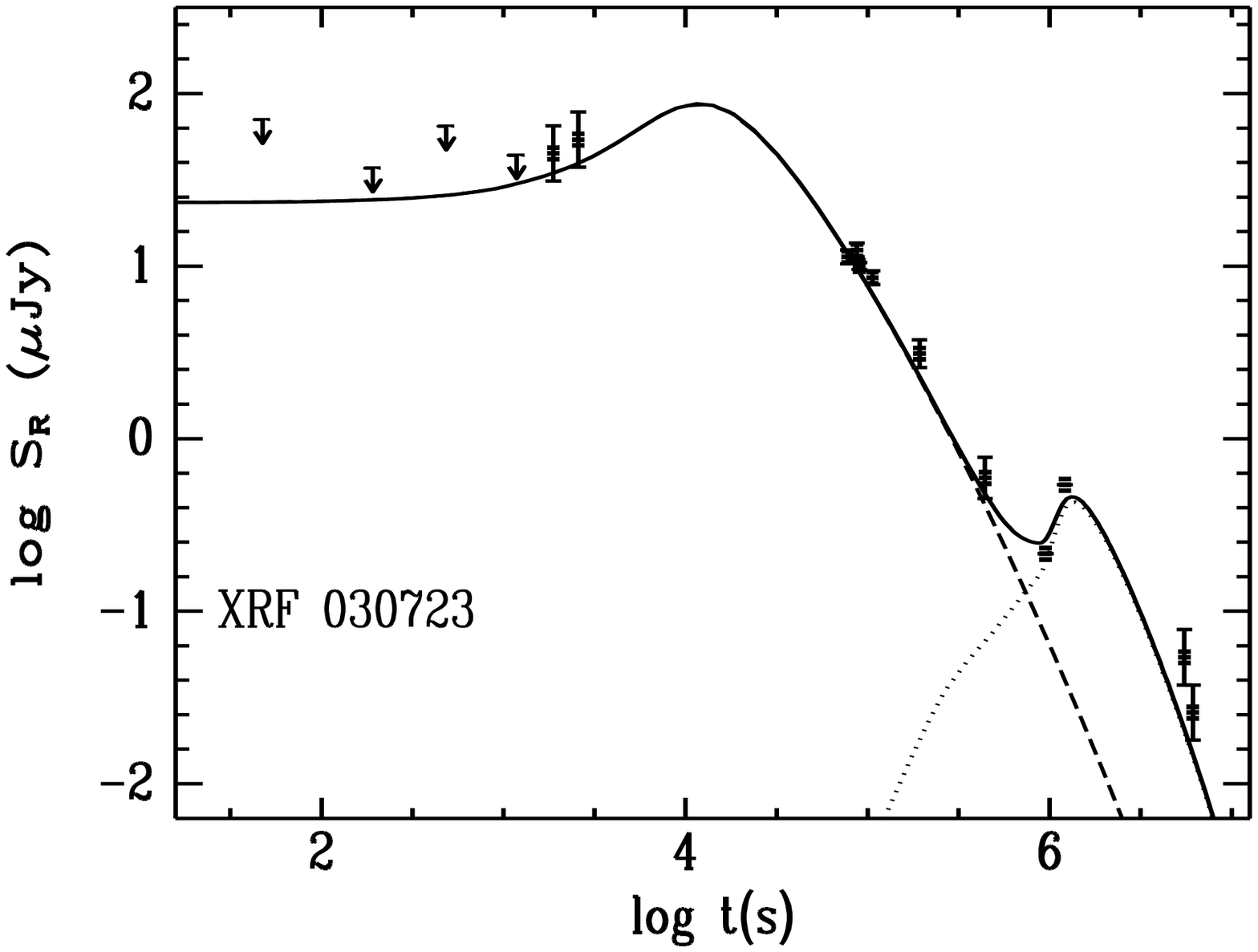, angle=-90, height=110mm, width=12.5cm, 
bbllx=120pt, bblly=125pt, bburx=530pt, bbury=575pt}
\caption{Fit to the optical afterglow of XRF 030723 by using the two-component
  jet model. Parameter values adopted in this plot are given in section 4 of
  the main text. The dashed line corresponds to emission from the wide 
  component and the dotted line is for the narrow component. The solid line
  illustrates the total light curve. Observed data points have been taken from
  the GCN (Prigozhin et al. 2003; Fox et al. 2003; Dullighan et al. 2003a, b; 
  Bond 2003; Smith et al. 2003; Fynbo et al. 2003b, c; Kawai et al. 2003). 
  Note that the error bars of the two points near $10^6$ s are not plotted, 
  since they are not available in the literature.   } 
\label{fig6}
\end{figure}


\clearpage

\begin{thebibliography}{}
\bibitem[]{694} Andersen, M., et al. 1999, Science, 283, 2075 
\bibitem[]{695} Barraud, C., et al. 2003, A\&A, 400, 1021 
\bibitem[]{696} Berger, E., Kulkarni, S. R., \& Frail, D. A. 2003a, ApJ, 590, 379 
\bibitem[]{697} Berger, E., et al. 2003b, Nature, 426, 154 
\bibitem[]{698} Bloom, J. S., et al. 1998, ApJ, 508, L21 
\bibitem[]{699} Bloom, J. S., Frail, D., \& Kulkarni, S. R. 2003, ApJ, 594, 674  
\bibitem[]{700} Bond, H. E. 2003, GCN, 2329 
\bibitem[]{701} Butler, N., et al. 2003a, GCN, 2328 
\bibitem[]{702} Butler, N., et al. 2003b, GCN, 2347 
\bibitem[]{703} Coburn, W., \& Boggs, S. E. 2003, Nature, 423, 415 
\bibitem[]{704} Dai, Z. G., \& Gou, L. J. 2001, ApJ, 552, 72 
\bibitem[]{705} Dai, Z. G., \& Lu, T. 2002, ApJ, 565, L87 
\bibitem[]{706} Dai, Z. G., \& Wu, X. F. 2003, ApJ, 591, L21
\bibitem[]{707} Dado, S., Dar, A., \& De R\'ujula, A., 2003, astro-ph/0309294
\bibitem[]{708} Dermer, C. D., Chiang, J., \& B\"ottcher, M. 1999, ApJ, 513, 656 
\bibitem[]{709} Dullighan, A., Butler, N., Vanderspek, R., Villasenor, J., \& Ricker, G.
    2003a, GCN, 2326 
\bibitem[]{711} Dullighan, A., Butler, N., Vanderspek, R., Villasenor, J., \& Ricker, G.
    2003b, GCN, 2336 
\bibitem[]{713} Fox, D. B., et al. 2003, GCN, 2323 
\bibitem[]{714} Frail, D. A., et al. 2000, ApJ, 538, L129
\bibitem[]{715} Frail, D. A., et al. 2001, ApJ, 562, L55 
\bibitem[]{716} Frontera, F., et al. 2000, ApJ, 540, 697
\bibitem[]{717} Fynbo, J. P. U., et al. 2003a, GCN, 2327 
\bibitem[]{718} Fynbo, J. P. U., et al. 2003b, GCN, 2345 
\bibitem[]{719} Fynbo, J. P. U., et al. 2003c, GCN, 2403 
\bibitem[]{720} Granot, J., \& Kumar, P. 2003, ApJ, 591, 1086
\bibitem[]{721} Granot, J., Nakar, E., \& Piran, T. 2003, Nature, 426, 138
\bibitem[]{722} Granot, J., Panaitescu, A., Kumar, P., \& Woosley, S. E. 2002, 
    ApJ, 570, L61 
\bibitem[]{724} Gruzinov, A. 1999, ApJ, 525, L29 
\bibitem[]{725} Harrison, F. A., et al. 1999, ApJ, 523, L121 
\bibitem[]{726} Heise, J., in't Zand, J. J. M., Kippen, R. M., \& Woods, P. M. 2003, AIP 
    Conf. Proc., 662, 229                   
\bibitem[]{728} Hjorth, J., et al. 1999, Science, 283, 2073 
\bibitem[]{729} Huang, Y. F., \& Cheng, K. S. 2003, MNRAS, 341, 263 
\bibitem[]{730} Huang, Y. F., Dai, Z. G., \& Lu, T. 1999a, Chin. Phys. Lett., 16, 775 
\bibitem[]{731} Huang, Y. F., Dai, Z. G., \& Lu, T. 1999b, MNRAS, 309, 513 
\bibitem[]{732} Huang, Y. F., Dai, Z. G., \& Lu, T. 2002, MNRAS, 332, 735 
\bibitem[]{733} Huang, Y. F., Dai, Z. G., \& Lu, T. 2000a, MNRAS, 316, 943 
\bibitem[]{734} Huang, Y. F., Gou, L. J., Dai, Z. G., \& Lu, T. 2000b, ApJ, 543, 90 
\bibitem[]{735} Jin, Z. P., \& Wei, D. M. 2003, A\&A, submitted (astro-ph/0308061) 
\bibitem[]{736} Kawai, N., et al. 2003, GCN, 2412
\bibitem[]{737} Kippen, R. M., Woods, P. M., Heise, J., in't Zand, J. J. M., Briggs, 
    M. S., \& Preece, R. D. 2003, AIP Conf. Proc., 662, 244  
\bibitem[]{739} Kulkarni, S. R., et al. 1998, Nature, 393, 35
\bibitem[]{740} Kulkarni, S. R., et al. 1999, Nature, 398, 389 
\bibitem[]{741} Kumar, P., \& Granot, J. 2003, ApJ, 591, 1075 
\bibitem[]{742} Lazzati, D., Rossi, E., Covino, S., Ghisellini, G., \& Malesani, D.
    2002, A\&A, 396, L5
\bibitem[]{744} M\'esz\'aros, P., Rees, M. J., \& Wijers, R. A. M. J. 1998, ApJ, 499, 301 
\bibitem[]{745} Mitra, A. 2000, A\&A, 359, 413 
\bibitem[]{746} Nakamura, T. 1999, ApJ, 522, L101 
\bibitem[]{747} Nakar, E., Piran, T., \& Granot, J. 2003, New Astron, 8, 495
\bibitem[]{748} Panaitescu, A., \& M\'esz\'aros, P. 1998, ApJ, 493, L31 
\bibitem[]{749} Panaitescu, A., \& Kumar, P. 2001, ApJ, 560, L49 
\bibitem[]{750} Prigozhin, G., et al. 2003, GCN, 2313 
\bibitem[]{751} Ramirez-Ruiz, E., Celotti, A., Rees, M. J. 2002, MNRAS, 337, 1349
\bibitem[]{752} Rhoads, J. E. 1997, ApJ, 487, L1 
\bibitem[]{753} Rhoads, J. E. 1999, ApJ, 525, 737 
\bibitem[]{754} Rol, E., et al. 2003, A\&A, 405, L23 
\bibitem[]{755} Rossi, E., Lazzati, D., \& Rees, M. J. 2002, MNRAS, 332, 945 
\bibitem[]{756} Sakamoto, T., et al. 2003, ApJ, accepted (astro-ph/0309455)
\bibitem[]{757} Salmonson, J. D. 2003, ApJ, 592, 1002 
\bibitem[]{758} Sari, R. 1998, ApJ, 494, L49 
\bibitem[]{759} Sari, R., Piran, T., \& Halpern, J. P. 1999, ApJ, 519, L17  
\bibitem[]{760} Sari, R., Piran, T., \& Narayan, R. 1998, ApJ, 497, L17 
\bibitem[]{761} Sheth, K., Frail, D. A., White, S., Das, M., Bertoldi, F., Walter, F., 
    Kulkarni, S. R., \& Berger, E. 2003, ApJ, 595, L33 
\bibitem[]{763} Smith, D. A., et al. 2003, GCN, 2338 
\bibitem[]{764} Soderberg, A. M., et al. 2002, GCN, 1554
\bibitem[]{765} Soderberg, A. M., et al. 2003a, GCN, 2330 
\bibitem[]{766} Soderberg, A. M., et al. 2003b, ApJ, submitted (astro-ph/0311050) 
\bibitem[]{767} Waxman, E. 1997, ApJ, 491, L19 
\bibitem[]{768} Waxman, E. 2003, Nature, 423, 388
\bibitem[]{769} Woods, E., \& Loeb, A. 1999, ApJ, 523, 187 
\bibitem[]{770} Wu, X. F., Dai, Z. G., Huang, Y. F., \& Lu, T. 2003, MNRAS, 342, 1131 
\bibitem[]{771} Yamazaki, R., Ioka, K., \& Nakamura, T. 2002, ApJ, 571, L31 
\bibitem[]{772} Yamazaki, R., Ioka, K., \& Nakamura, T. 2003, ApJ, 593, 941 
\bibitem[]{773} Zhang, B., Dai, X., Lloyd-Ronning, N. M., \& M\'esz\'aros, P. 2003, 
    ApJ, submitted (astro-ph/0311190)
\bibitem[]{775} Zhang, B., \& M\'esz\'aros, P. 2002a, ApJ, 571, 876
\bibitem[]{776} Zhang, B., \& M\'esz\'aros, P. 2002b, ApJ, 581, 1236
\bibitem[]{777} Zhang, W., Woosley, S. E., \& Heger, A. 2003a, ApJ, 
    submitted (astro-ph/0308389) 
\bibitem[]{779} Zhang, W., Woosley, S. E., \& MacFadyen, A. I. 2003b, ApJ, 586, 356 
\end{thebibliography}
\end{document}